# A Rocky Composition for an Earth-sized Exoplanet

Andrew W. Howard[1], Roberto Sanchis-Ojeda[2], Geoffrey W. Marcy[3],
John Asher Johnson[4], Joshua N. Winn[2], Howard Isaacson[3], Debra A. Fischer[5],
Benjamin J. Fulton[1], Evan Sinukoff[1], Jonathan J. Fortney[6]

[1]Institute for Astronomy, University of Hawaii at Manoa, 2680 Woodlawn Drive, Honolulu, HI 96822 USA
[2]Department of Physics, and Kavli Institute for Astrophysics and Space Research, Massachusetts Institute of Technology, Cambridge, MA, 02139 USA
[3]Astronomy Department, University of California, Berkeley, CA 94720 USA
[4]Harvard-Smithsonian Center for Astrophysics, 60 Garden Street, Cambridge, MA 02138 USA
[5]Department of Astronomy, Yale University, New Haven, CT 06510 USA
[6]Department of Astronomy and Astrophysics, University of California, Santa Cruz, CA 95064 USA

**Planets with sizes between that of Earth (with radius $R_\oplus$) and Neptune (about 4 $R_\oplus$) are now known to be common around Sun-like stars[1,2,3]. Most such planets have been discovered through the transit technique, by which the planet's size can be determined from the fraction of starlight blocked by the planet as it passes in front of its star. Measuring the planet's mass—and hence its density, which is a clue to its composition—is more difficult. Planets of size 2-4 $R_\oplus$ have proven to have a wide range of densities, implying a diversity of compositions[4,5], but these measurements did not extend down to planets as small as Earth. Here we report Doppler spectroscopic measurements of the mass of the Earth-sized planet Kepler-78b, which orbits its host star every 8.5 hours (ref. 6). Given a radius of 1.20 ± 0.09 $R_\oplus$ and mass of 1.69 ± 0.41 $M_\oplus$, the planet's mean density of 5.3 ± 1.8 g cm$^{-3}$ is similar to the Earth's, suggesting a composition of rock and iron.**

Kepler-78 is one of approximately 150,000 stars whose brightness was precisely measured at 30-minute intervals for four years by the NASA Kepler spacecraft[7]. This star is somewhat smaller, less massive, and younger than the Sun (Table 1). Every 8.5 hours the star's brightness declines by 0.02% as the planet Kepler-78b transits (passes in front of) the stellar disk. The planet's radius was originally measured[6] to be $1.16^{+0.19}_{-0.14}$ $R_\oplus$. The mass could not be measured, although masses >8 $M_\oplus$ could be ruled out because the planet's gravity would have deformed the star and produced brightness variations that were not detected.

We measured the mass of Kepler-78b by tracking the line-of-sight component of the host star's motion (the radial velocity, RV) due to the gravitational force of the planet. The RV analysis is challenging not only because the signal is expected to be small (~1-3 m s$^{-1}$) but also because the apparent Doppler shifts due to rotating starspots are much larger (~50 m s$^{-1}$ peak-to-peak). Nevertheless the detection proved to be possible, thanks to the precisely known orbital period and phase of Kepler-78b that cleanly separated the timescale of spot variations ($P_{\rm rot} \approx 12.5$ days) from the much shorter timescale of the planetary orbit ($P \approx 8.5$ hours). We adopted a strategy of intensive Doppler measurements spanning 6-8 hours per night, long enough to cover nearly the entire orbit and short enough for the spot variations to be nearly frozen out.

We measured RVs using optical spectra of Kepler-78 that we obtained from the High Resolution Echelle Spectrometer (HIRES)[8] on the 10-m Keck I Telescope. These Doppler shifts were computed relative to a template spectrum with a standard algorithm[9] that uses a spectrum of molecular iodine superposed on the stellar spectrum as a reference for the wavelength scale and instrumental profile of HIRES (Supplementary Table 1). Exposures lasted 15-30 minutes depending on conditions and produced RVs with 1.5-2.0 m s$^{-1}$ uncertainties. The time series of RVs spans 38 days, with large velocity offsets between nights due to spots (Fig. 1). Within each night the RVs vary by typically 2-4 m s$^{-1}$ and show coherence on shorter time scales.

We modeled the RV time series as the sum of two components. One component was a sinusoidal function representing orbital motion (assumed to be circular). The orbital period and phase were held fixed at the photometrically-determined values; the only free parameters were the Doppler amplitude $K$, an arbitrary RV zero point, and a velocity 'jitter' term $\sigma_{jitter}$ to account for additional RV noise. The second component of the model, representing the spot variations, was the sum of three sinusoidal functions with periods $P_{rot}$, $P_{rot}/2$, and $P_{rot}/3$. The amplitudes and phases of the sinusoids and $P_{rot}$ were free parameters. All together there were 10 parameters and 77 data points. Using a Markov Chain Monte Carlo method to sample the allowed combinations of the model parameters, we found $K = 1.66 \pm 0.40$ m s$^{-1}$, corresponding to $M_{pl} = 1.69 \pm 0.41\ M_{\oplus}$ (Fig. 1). This planet mass is consistent with an independent measurement using the HARPS-N spectrometer[10].

Several tests were performed to gauge the robustness of the spot model. First, we varied the number of harmonics, checking at each stage whether any improvement in the fit was statistically significant. The three-term model was found to provide significant improvement over one-term and two-term models, but additional harmonics beyond $P_{rot}/3$ did not provide significant improvement. Second, we used a different spot model in which the spot-induced variation was taken to be a linear function of time specific to each night. The constant and slope of each nightly function were free parameters. With this model we found $M_{pl} = 1.50 \pm 0.44\ M_{\oplus}$, consistent with the preceding results (see Methods and Extended Data Figure 3). The larger uncertainty can be attributed to the greater flexibility of this piecewise-linear spot model, which permits discontinuous and probably unphysical variations between consecutive nights.

Kepler-78b is now the smallest exoplanet for which both the mass and radius are known accurately (Fig. 2), extending the domain of such measurements into the neighborhood of Earth and Venus. Kepler-78b is 20% larger than Earth and is 69% more massive, suggesting commonality with the other low-mass planets (4-8 $M_{\oplus}$) below the rock composition contour in Fig. 2b. They are all consistent with rock/iron compositions and negligible atmospheres.

We explored some possibilities for the interior structure of Kepler-78b using a simplified two-component model[11] consisting of an iron core surrounded by a silicate mantle ($Mg_2SiO_4$). This model correctly reproduces the masses of Earth and Venus given their radii and assuming a composition of 67% silicate rock and 33% iron by mass. Applied to Kepler-78b, the model gives an iron fraction of 20% ± 33%, similar to that of Earth and Venus but smaller than that of Mercury (≈60%, [12]).

With a star-planet separation of 0.01 astronomical units (the Earth-Sun distance, AU), the dayside of Kepler-78b is heated to a temperature of 2300-3100 K. Any gaseous atmosphere around Kepler-78b would probably have been long ago lost to photoevaporation by the intense starlight[13]. However, based on the measured surface gravity of 11 m s$^{-2}$, the liquid and solid portions of the planet should be stable against mass loss of the sort[14] that is apparently destroying the smaller planet KIC 12557548b[15].

Kelper-78b is a member of an emerging class of planets with orbital periods of less than half a day[6,16,17]. Another member is KOI 1843.03 (refs. 18,19), which has been shown to have a high density (≳7 g cm$^{-3}$), although the deduction in that case was based on the theoretical requirement to avoid tidal destruction rather than direct measurement. That planet's minimum density is similar to our estimated density for Kepler-78b ($5.3^{+2.0}_{-1.6}$ g cm$^{-3}$). These two planets provide a stark contrast to Kepler-11f, which has a similar mass to Kepler-78b, but a density that is 10 times smaller[20].

With only a handful of low-mass planets with measured densities known (Fig. 2b), we see solid planets primarily in highly irradiated, close-in orbits and low-density planets swollen by thick atmospheres in somewhat cooler orbits. Measurements of additional planet masses and radii are needed to assess the significance of this pattern. Additional ultrashort period planets with detectable Doppler amplitudes ($K \propto P^{-1/3}$) have been identified by the Kepler mission and are ripe for mass measurements. With an ensemble of future measurements, the masses and radii of ultrashort period planets may reveal a commonality or diversity of density and composition. This knowledge of hot solid planets may be relevant to problems such as the interiors of cooler extrasolar planets with atmospheres, the range of core sizes in giant planet formation, and Mercury's unusually high iron abundance.

**Methods Summary:** We fit Keck-HIRES spectra of Kepler-78 with stellar atmosphere models using the Spectroscopy Made Easy software package to measure the star's temperature, gravity, and iron abundance. These spectroscopic parameters were used to estimate the host star's mass, radius, and density—crucial parameters to determine the planet's mass, radius, and density—from empirical relationships calibrated by precisely characterized binary star systems. Using this stellar density as a constraint, we reanalyzed the Kepler photometry to refine the planet radius measurement. We observed Kepler-78 with HIRES using standard procedures including sky spectrum subtraction and wavelength calibration with a reference iodine spectrum. We measured high-precision relative RVs using a forward model where the de-convolved stellar spectrum is Doppler shifted, multiplied by the normalized high-resolution iodine transmission spectrum, convolved with an instrumental profile, and matched to the observed spectra using a Levenberg-Marquardt algorithm that minimizes the $\chi^2$ statistic. The time series RVs on eight nights were analyzed with several parametric models to account for the small-amplitude, periodic signal from the orbiting planet and the larger amplitude, quasi-periodic apparent Doppler shifts due to rotating starspots. In our adopted harmonic spot model the starspot signal was modeled as a sum of sine functions whose amplitudes and phases were free parameters. We sampled the multi-dimensional model parameter space with a Markov Chain Monte Carlo algorithm to estimate parameter confidence intervals and to account for covariance between parameters. We found multiple families of models that add described the data well and they gave consistent measures of the Doppler amplitude, which is proportional to the mass of Kepler-78b.

**Supplementary Information** is linked to the online version of the paper at www.nature.com/nature.

**Acknowledgments** This Letter and a companion by F. Pepe and colleagues[10] were submitted simultaneously and are the result of a coordinated, independent RV observations and analyses of Kepler-78. We thank the HARPS-N team for their collegiality. We also thank E. Chiang, I. Crossfield, R. Kolbl, E. Petigura, and D. Huber for discussions, S. Howard for support, C. Dressing for a convenient packaging of stellar models, and A. Hatzes for a thorough review. This work was based on observations at the W. M. Keck Observatory granted by the University of Hawaii, the University of California, and the California Institute of Technology. We thank the observers who contributed to the measurements reported here and acknowledge the efforts of the Keck Observatory staff. We extend special thanks to those of Hawaiian ancestry on whose sacred mountain of Mauna Kea we are privileged to be guests. Kepler was competitively selected as the tenth Discovery mission with funding provided by NASA's Science Mission Directorate. J.N.W. and R.S.O. acknowledge support from the Kepler Participating Scientist program. A.W.H. acknowledges funding from NASA grant NNX12AJ23G.

**Author Contributions** This measurement was conceived and planned by A.W.H., G.W.M., J.A.J., J.N.W., and R.S.O. HIRES observations were conducted by A.W.H., G.W.M., H.I, B.J.F., and E.S. The HIRES spectra were reduced and Doppler analyzed by A.W.H., G.W.M., H.I., and J.A.J. Data modeling was done primarily by A.W.H. and R.S.O. A.W.H. was the primary author of the manuscript, with important contributions from J.N.W., R.S.O., and J.J.F. Figures were generated by A.W.H., R.S.O, B.J.F., and E.S. All authors discussed the results, commented on the manuscript, and contributed to the interpretation.

**Author Information** Reprints and permissions information is available at www.nature.com/reprints. The authors declare no competing financial interest. Readers are welcome to comment on the online version of this article at www.nature.com/nature. Correspondence and requests for materials should be addressed to A.W.H. (howard@ifa.hawaii.edu) or R.S.-O. (rsanchis86@gmail.com).

**Table 1 | Kepler-78 System Properties.**

| Stellar Properties | |
|---|---|
| Names | Kepler-78, KIC 8435766, Tycho 3147-188-1 |
| Effective Temperature, $T_{eff}$ | 5121 ± 44 K |
| Logarithm of surface gravity, log $g$ | 4.61 ± 0.06 ($g$ in cm s$^{-2}$) |
| Iron abundance, [Fe/H] | -0.08 ± 0.04 dex |
| Projected rotational velocity, $V\sin i$ | 2.6 ± 0.5 km s$^{-1}$ |
| Mass, $M_{star}$ | 0.83 ± 0.05 $M_{sun}$ |
| Radius, $R_{star}$ | 0.74 ± 0.05 $R_{sun}$ |
| Density, $\rho_{star}$ | $2.8^{+0.7}_{-0.6}$ g cm$^{-3}$ |
| Age | 625 ± 150 million years |
| **Planetary Properties** | |
| Name | Kepler-78b |
| Mass, $M_{pl}$ | 1.69 ± 0.41 $M_\oplus$ |
| Radius, $R_{pl}$ | 1.20 ± 0.09 $R_\oplus$ |
| Density, $\rho_{pl}$ | $5.3^{+2.0}_{-1.6}$ g cm$^{-3}$ |
| Surface gravity, $g_{pl}$ | $11.4^{+3.5}_{-3.1}$ m s$^{-2}$ |
| Iron fraction | 0.20 ± 0.33 (two component rock/iron model) |
| Orbital period, $P_{orb}$ (from [6]) | 0.35500744 ± 0.00000006 days |
| Transit epoch, $t_c$ (from [6]) | 2454953.95995 ± 0.00015 (BJD$_{TBD}$) |
| **Additional Parameters** | |
| $(R_{pl}/R_{star})^2$ | 217 ± 9 parts per million (ppm) |
| Scaled semi-major axis, $a/R_{star}$ | 2.7 ± 0.2 |
| Doppler amplitude, $K$ | 1.66 ± 0.40 m s$^{-1}$ |
| Systemic radial velocity | -3.59 ± 0.10 km s$^{-1}$ |
| Radial velocity jitter, $\sigma_{jitter}$ | 2.1 ± 0.3 m s$^{-1}$ |
| Radial velocity dispersion | 2.6 m s$^{-1}$ (s.d. of residuals to best-fit model) |

The stellar effective temperature and iron abundance were obtained by fitting stellar-atmosphere models[21] to iodine-free HIRES spectra[9], subject to a constraint on the surface gravity based on stellar-evolutionary models[22]. We estimated the stellar mass and radius from empirically calibrated relationships between those spectroscopic parameters[23]. The refined stellar radius led to a refined planet radius. Planet mass and density were measured from the Doppler analysis. The stellar age is estimated from non-detection of lithium in the stellar atmosphere (Extended Data Figure 1), the stellar rotation period, and magnetic activity. See Methods for details. Parameter distributions are represented by median values and 68.3% confidence intervals. Correlations between transit parameters are shown in Extended Data Figure 2.

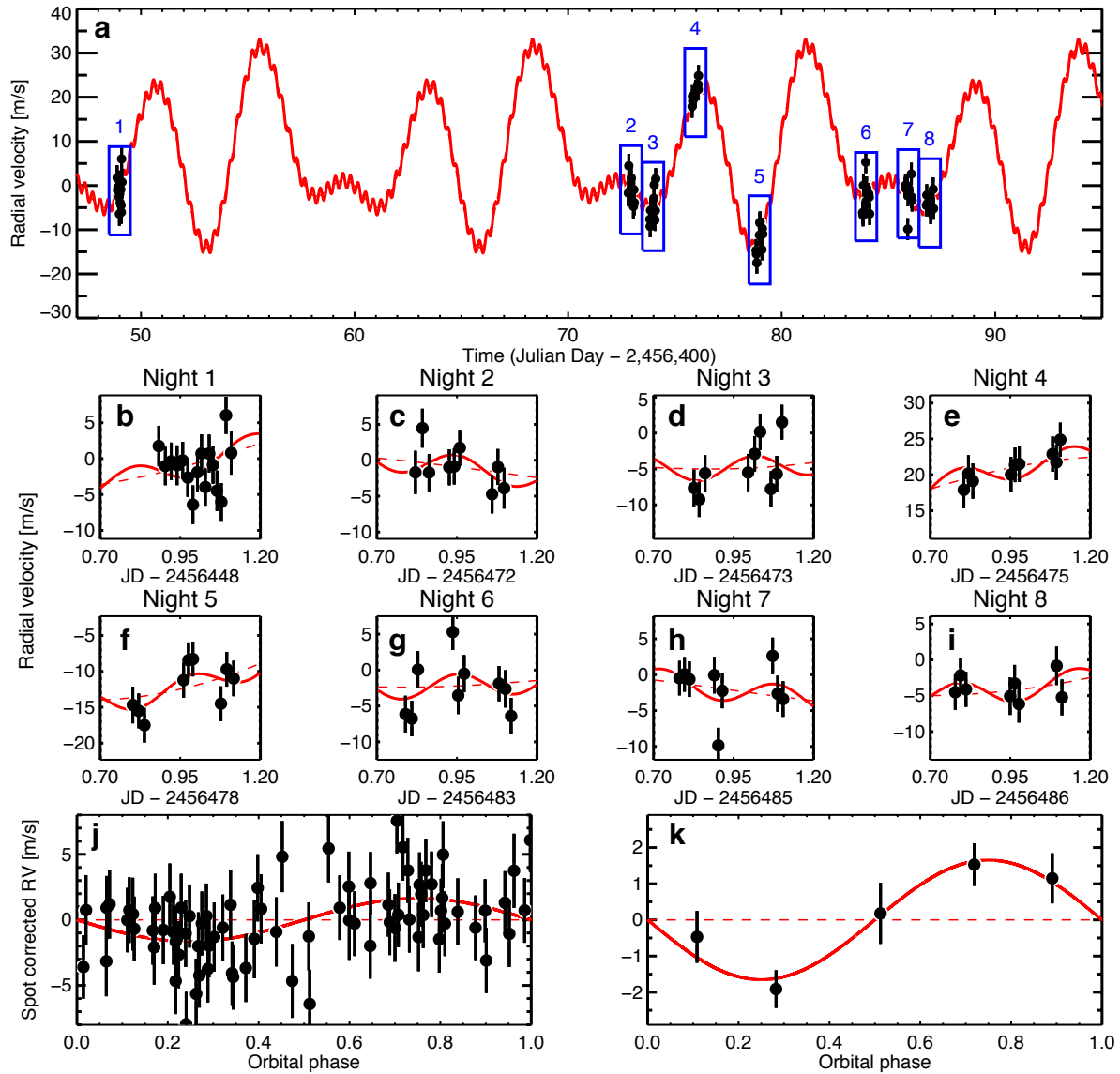

**Figure 1 | Apparent radial velocity (RV) variations of Kepler-78.** Panel **a** shows the 38-day time series of relative RVs (black filled circles) from Keck-HIRES along with the best-fitting model (red line), with short-term variations due to orbital motion and long-term variations due to rotating starspots. Blue boxes identify the eight nights when high-cadence measurements were undertaken. Panels **b-i** focus on those individual nights, showing the measured RVs (black filled circles), the spot+planet model (solid red lines), and spot model alone (dashed red lines). The lower panels show the phase-folded RVs after subtracting the best-fitting spot model (**j**), and after binning in orbital phase and computing the mean RVs and s.e.m. for error bars (**k**). Planetary transits occur at zero orbital phase. Each RV error bar in panels **a-j** represents the s.e.m. for the Doppler shifts of ~700 segments of a particular spectrum; it does not account for additional uncorrected RV "jitter" from astrophysical and instrumental sources.

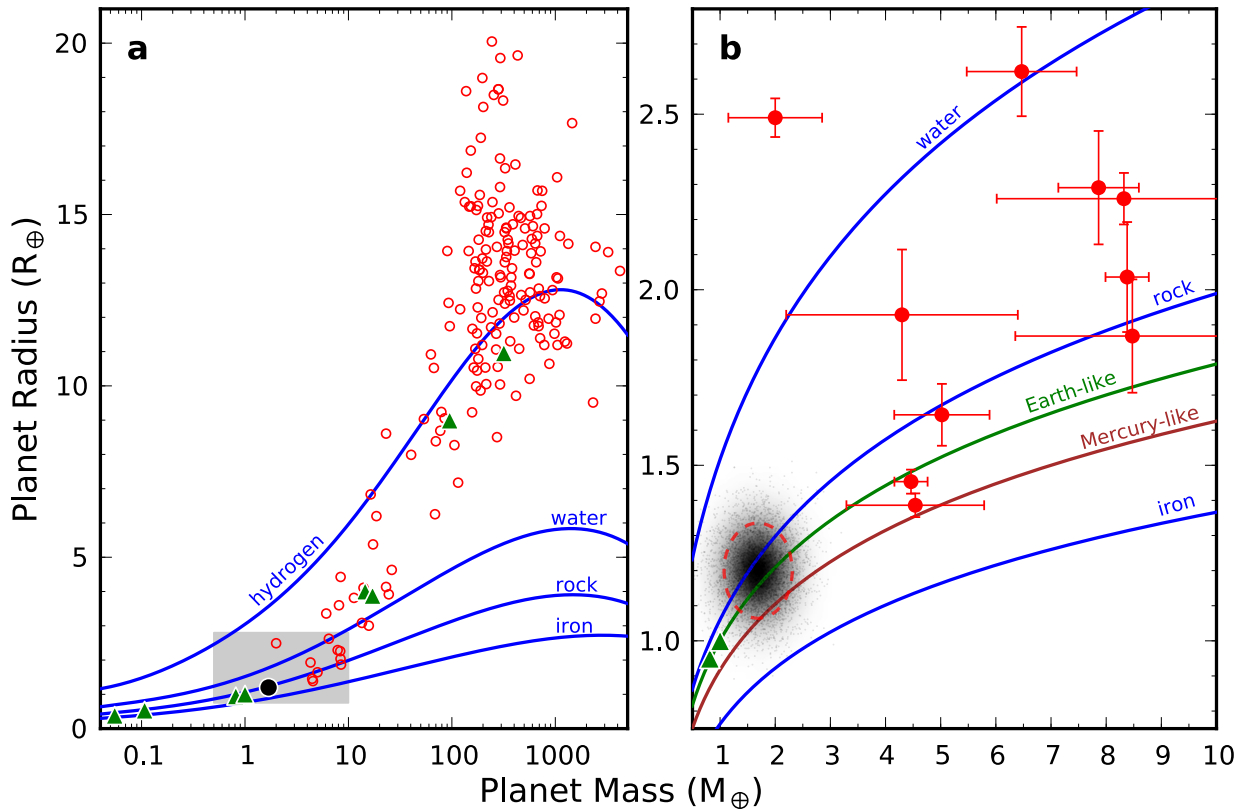

**Figure 2 | Masses and radii of well-characterized planets.** Extrasolar planets are denoted by red circles while Solar System planets are represented by green triangles. Panel **a** spans the full range of sizes and masses on logarithmic axes. The shaded gray rectangle denotes the range of parameters shown in panel **b** on linear mass and radius axes. Kepler-78b is depicted as a black filled circle (**a**) and as a distribution of allowed masses and radii with a red ellipse marking the 68% confidence region (**b**). Model mass-radius relationships[24,11] for idealized planets consisting of pure hydrogen, water, rock ($Mg_2SiO_4$), and iron are shown as blue lines. Green and brown lines denote Earth-like composition (67% rock, 33% iron) and Mercury-like composition (40% rock, 60% iron). Exoplanet masses, radii, and their associated errors are from the Exoplanet Orbit Database[25] (http://exoplanets.org; downloaded on 1 September 2013). Planets with fractional mass uncertainties of > 50% are not shown.

**Methods:**

**Stellar Characterization.** We fit three Keck-HIRES spectra of Kepler-78 with stellar atmosphere models using Spectroscopy Made Easy (SME[26]). The spectra have per-pixel signal-to-noise ratios of 220 at 550 nm. We used the standard wavelength intervals, line data, and methodology[27]. Kepler-78 does not have a measured parallax to constrain luminosity and gravity. The initial analysis gave an effective temperature $T_{eff}$ = 5119 ± 44 K, gravity log $g$ = 4.751 ± 0.060 (cgs), iron abundance [Fe/H] = -0.054 ± 0.040 dex, and projected rotational velocity $V\sin i$ = 2.2 ± 0.5 km s$^{-1}$. These values are the mean of the SME results for the three spectra and the error bars are limited by systematics[27]. Because this combination of $T_{eff}$ and log $g$ is inconsistent with the Dartmouth stellar evolutionary model[21], we recomputed stellar parameters with log $g$ fixed at the value model-predicted by a stellar model at the value of $T_{eff}$ from SME giving the stellar parameters in Table 1. Note that adopted $V\sin i$ = 2.6 ± 0.5 km s$^{-1}$ is consistent with an expectation based on a stellar rotation, size, and an equatorial viewing geometry: $V\sin i \approx V_{rot} \approx 2\pi R_{star}/P_{rot}$ = 3.0 km s$^{-1}$.

We estimated the stellar mass and radius using empirical relationships[23] based on non-interacting binary systems that parameterize $R_{star}$ and $M_{star}$ as functions of log $g$, $T_{eff}$, and [Fe/H]. We propagated the errors on the three SME-derived inputs to obtain $R_{star}$ = 0.74 ± 0.05 $R_{sun}$ and $M_{star}$ = 0.83 ± 0.05 $M_{sun}$. The $M_{star}$ uncertainty comes from the 6% fractional scatter in the mass-radius relationship[23]. We adopt these values when computing $R_{pl}$ and $M_{pl}$. We checked for self-consistency of the empirical calibration by computing log $g$ from the derived $R_{star}$ and $M_{star}$, giving log $g$ = 4.62 ± 0.06 (cgs).

As a consistency check, we explored two additional estimates of stellar parameters. First, the mass and radius from an evolutionary track in the Dartmouth model (1 Gyr, [m/H] = 0) that match our adopted $T_{eff}$ and log $g$ values are $R_{star}$ = 0.77 ± 0.04 $R_{sun}$ and $M_{star}$ = 0.85 ± 0.05 $M_{sun}$. These values are consistent with our adopted results. Second, we used a recent study[28] of stellar angular diameters that parameterized $R_{star}$ as a function of $T_{eff}$. This gives $R_{star}$ = 0.77 ± 0.03 $R_{sun}$, where the uncertainty is the median absolute deviation on the calibration star radii.

We note that Kepler-78 has remarkably similar properties to the transiting planet host star HD 189733. These properties[29,30] include $T_{eff}$ = 5040 K, log $g$ = 4.587 (cgs), $R_{star}$ = 0.76 $R_{sun}$ and $M_{star}$ = 0.81 $M_{sun}$, log $R'_{HK}$ = -4.50, and $P_{rot}$ = 11.9 days. HD 189733 has spot-induced RV variations[31] of ~15 m s$^{-1}$ (rms).

The rotation period of Kepler-78 was previously measured to be 12.5 ± 1.0 days[6]. Using a relationship[31] between age, mass, and rotation period, we estimate an age of 750 ± 150 Myr. The stellar age can also be estimated from the stellar magnetic activity measured by the $S_{HK}$ index. We computed the spectral-type-independent activity index, log $R'_{HK}$, for all HIRES observations of this star and found a median value of -4.52 with a 1-σ range of ±0.03. The computation made use of an estimated $B-V = 0.873$, converted[9] from $T_{eff}$. This level of activity is consistent with the value for stars in the 625 Myr old Hyades cluster[32]. We also constrained the age by searching for the age-sensitive Li I absorption line at 6708 Å. Lithium is depleted relatively quickly in stars of this spectral type owing to convective mixing. Based on Li I measurements in three clusters with known ages[33], our non-detection (Extended Data Figure 1) suggests an age greater than ~500 Myr. These three ages are self-consistent. We adopt an age of 625 Myr with an approximate age uncertainty of 150 Myr. We expect a star of this age and activity to have spots that cause RV variations at the > 10 m s$^{-1}$ level.

**Transit analysis.** Transit parameters are crucial to estimate the planet radius, which in turn affects our ability to estimate the composition of the planet. These parameters were measured previously with the discovery of Kepler-78b[6]. In that study the impact parameter $b$ was nearly unconstrained because the 30 min average of the Kepler long-cadence data cannot resolve the transit ingress time. This leads to an increased uncertainty on transit depth due to the stellar limb darkening profile. We constrained the transit parameters using the stellar density ($\rho_{star}$) obtained from the spectroscopic analysis. Assuming a circular orbit,

$$\rho_{star} = (3\pi/GP^2)(a/R_{star})^3,$$

where $a/R_{star}$ is the scaled semi-major axis[34]. This gives $a/R_{star} = 2.7 \pm 0.2$, a much tighter constraint than from the transit light curve alone ($a/R_{star} = 3.0^{+0.5}_{-1.0}$).

Aside from this additional constraint, our transit analysis is similar to the one in [6]. In brief, we analyzed the Kepler long-cadence data from Q1 through Q15 (a total of 3.7 years of nearly continuous observations) to construct a filtered, phase-folded light curve with a final cadence of 2 minutes. The light curve is modeled with a combination of a transit model[35], a model for the out-of-transit modulations, and an occultation model. The most relevant transit parameters are the impact parameter, the ratio of stellar radius to orbital distance, and the zero-limb-darkening transit depth. The model is calculated with a cadence of 15 seconds and averaged over the 30 minute cadence of Kepler. In this new analysis, $R_{star}/a$ is subjected to a Gaussian prior ($2.7 \pm 0.2$), which leads to a well-measured impact parameter and a reduced uncertainty for the transit depth. We found the best-fit solution and explored parameter space using a Markov Chain Monte Carlo (MCMC) algorithm. The final parameters are $(R_{pl}/R_{star})^2 = 217^{+9}_{-10}$, $b = 0.68^{+0.05}_{-0.08}$, orbital inclination $i = 75.2^{+2.6}_{-2.1}$ deg, transit duration = $0.813 \pm 0.014$ hours, and $R_{pl} = 1.20 \pm 0.09$ $R_\oplus$. Error bars encompass 68.3% confidence intervals. Parameter correlations are plotted in Extended Data Figure 2. These values are compatible with the previous estimate that was not constrained by $\rho_{star}$ [6].

**Radial Velocity Measurements.** We observed Kepler-78 with the HIRES echelle spectrometer[8] on the 10-m Keck I telescope using standard procedures. Observations were made with the C2 decker (14 x 0.86 arcseconds). This slit is long enough to simultaneously record spectra of Kepler-78 and the faint night sky. We subtracted the sky spectra from the spectra of Kepler-78 during the spectral reduction[36].

Light from the telescope passed through a glass cell of molecular iodine cell heated to 50° C. The dense set of molecular absorption lines imprinted on the stellar spectra in 5000-6200 Å provide a robust wavelength scale against which Doppler shifts are measured, as well as strong constraints on the instrumental profile at the time of each observation[37,38]. We also obtained three iodine-free "template" spectra of Kepler-78 using the B3 decker (14 x 0.57 arcseconds). These spectra were used to measure stellar parameters, as described above. One of them was de-convolved using the instrumental profile measured from spectra of rapidly rotating B stars observed immediately before and after. This de-convolved spectrum served as a "template" for the Doppler analysis.

The HIRES observations span 38 days. On eight nights we observed Kepler-78 intensively, covering 6-8 hours per night. We also gathered a single spectrum on six additional nights to monitor the RV variations from spots. These once-per-night RVs were not used to determine the planetary mass and are shown in Extended Data Figure 3 but not in Figure 1.

We measured high-precision relative RVs using a forward model where the de-convolved stellar spectrum is Doppler shifted, multiplied by the normalized high-resolution iodine transmission spectrum, convolved with an instrumental profile, and matched to the observed spectra using a Levenberg-Marquardt algorithm that minimizes the $\chi^2$ statistic[9]. In this algorithm, the RV is varied (along with nuisance parameters describing the wavelength scale and instrumental profile) until the $\chi^2$ minimum is reached.

The times of observation (in heliocentric Julian days, HJD), RVs relative to an arbitrary zero point, and error estimates are listed in Supplementary Table 1 and plotted in Extended Data Figure 3. Each RV error is the standard error on the mean RV of ~700 spectral chunks (each spanning ~2 Å) that are separately Doppler analyzed. These error estimates do not account for systematic Doppler shifts from instrumental or stellar effects. We also measured the $S_{HK}$ index for each HIRES spectrum. This index measures the strength of the inversion cores of the Ca II H & K absorption lines and correlates with stellar magnetic activity[39].

We measured the absolute RV of Kepler-78 relative to the Solar System barycenter using telluric sky lines as a reference[40]. The distribution of telluric RVs has a median value of -3.59 km s$^{-1}$ and a standard deviation of 0.10 k s$^{-1}$.

**Harmonic RV Spot Model.** Kepler-78 is young active star, as demonstrated by the large stellar flux variations observed with Kepler. A previous study[6] measured $P_{rot}$ = 12.5 ± 1.0 days using a

Lomb Scargle periodogram of the photometry. Inspection of the RVs measured over one month indeed show some repeatability with a timescale of ~12-13 days, a sign that starspots are also inducing a large RV signal (see Extended Data Figure 3). Based on previous work[41], we modeled the RV signal induced by spots with a primary sine function at the rotation period of the star, followed by a series of sine functions representing subharmonics of the stellar rotation. The planet-induced RV signal is modeled with a sinusoid, assuming zero eccentricity and using a linear ephemeris fixed to the best-fit orbital period and phase[6]. The final model for the RV at time $t$ is

$$RV(t) = -K \sin(2\pi (t-t_c)/P) + \gamma + \sum_i a_i \sin(\varphi_i + i\, 2\pi t / P_{rot})$$

where $K$ is the semi-amplitude of the planet-induced RV signal, $t_c$ is a time of transit, $P$ is the orbital period, i runs from 1 to $N$, where $N$ is the number of subharmonics used, $P_{rot}$ is the rotation period, and finally $a_i$ and $\varphi_i$ are the two parameters added for each of the $N$ subharmonics. The amplitude $a_i$ is always chosen to be positive, and $\varphi_i$ is constrained to be positive and smaller than $2\pi$. The time $t$ was set to zero at 2456446 in HJD format. $P$ and $t_c$ were held fixed to their photometrically determined values (Table 1). $K$ was free to take on positive and negative values to prevent a bias toward larger planet mass.

We used the Bayesian Information Criterion to choose the appropriate number of subharmonics. This criterion states that for each additional model parameter the standard $\chi^2$ function should decrease by at least $\ln(N_{meas})$ in order to be deemed statistically significant. In our case $N_{meas}$ = 77 (the number of RVs on the eight nights with intensive observations). For each subharmonic added, the best-fit $\chi^2$ should decrease by at least 8.7 for the more complex model to be justified. The best-fit $\chi^2$ values for $N$ = 1, 2, 3, 4, and 5 were 1822, 262, 163, 161, and 158, respectively. We used a Gaussian prior to control the rotation period in this analysis to select the number of harmonics. We adopted the $N$ = 3 model because adding additional subharmonics is not statistically justified.

This analysis does not rule out the possibility that non-consecutive subharmonics provide a better fit to the data. We checked that for a model with three subharmonics chosen from the first four, the first three subharmonics is the best combination. We also checked that choosing only two subharmonics out of the first four was never a better option than using the first three.

Using the spot model with the first three subharmonics, we used an MCMC algorithm to explore model parameter space. We added an RV "jitter" term $\sigma_{jitter}$ to account for the high value of the reduced $\chi^2$ in the best-fit model without jitter (a value of 2.1), following a standard procedure[42] to leave it as a free parameter. We maximized the logarithm of the likelihood function instead of minimizing the $\chi^2$ function. We estimate the parameters describing the planet be $K$ = 1.66 ± 0.40 m s$^{-1}$, $\gamma = 4.4^{+2.7}_{-2.0}$ m s$^{-1}$, and $\sigma_{jitter} = 2.08^{+0.32}_{-0.29}$ m s$^{-1}$. The parameters descripting the starspots are

$P_{rot} = 12.78 \pm 0.04$ days, $a_1 = 3.6^{+3.6}_{-1.4}$ m s$^{-1}$, $a_2 = 10.5^{+3.3}_{-2.4}$ m s$^{-1}$, $a_3 = 10.2^{+1.5}_{-1.3}$ m s$^{-1}$, $\varphi_1 = 4.4^{+1.2}_{-1.0}$, $\varphi_2 = 3.9^{+0.2}_{-0.3}$, and $\varphi_3 = 0.48 \pm 0.21$. These values are the median and 68.3% confidence regions of marginalized posterior distributions from the MCMC analysis. In this final run, $P_{rot}$ was not subject to a prior and yet the value is compatible with the photometric estimate. Our estimate of $K$ is inconsistent with zero at the 4-σ level. This 4-σ detection of Kepler-78b that is consistent with the orbital period and phase from Kepler gives us high confidence that we have detected the planet. The planet mass listed in Table 1 was computed from the values for $K$, $P$, $i$, and $M_{star}$ with the assumption of a circular orbit.

We searched for additional $\chi^2$ minima to assess the sensitivity of our mass measurement to the spot model. We found a second family of solutions with $K = 1.85 \pm 0.43$ m s$^{-1}$, $\sigma_{jitter} = 2.3 \pm 0.3$ m s$^{-1}$, and $P_{rot} = 12.3$ days. While our adopted solution with $P_{rot} = 12.8$ days is clearly preferred by the $\chi^2$ criterion ($\chi^2 = 163$ versus 180), the broader model search demonstrates that our mass determination is relatively insensitive to details of the spot model.

We also calculated the $K$ values with different combinations of two and three harmonics. For example, a model with the first, second, and fourth harmonic gives $K = 1.78 \pm 0.45$ m s$^{-1}$, with a slightly larger $\sigma_{jitter} = 2.3 \pm 0.3$ m s$^{-1}$. A model with the first four harmonics gives $K = 1.77 \pm 0.41$ m s$^{-1}$, with $\sigma_{jitter} = 2.2 \pm 0.3$ m s$^{-1}$. Second, we included all of the RVs (including the six RVs measured on nights without intensive observations) and fitted the complete data set with three and four consecutive harmonics, giving $K = 1.80 \pm 0.43$ m s$^{-1}$ and $\sigma_{jitter} = 2.4 \pm 0.3$ m s$^{-1}$ for the three harmonics model, and $K = 1.77 \pm 0.41$ m s$^{-1}$ and $\sigma_{jitter} = 2.2 \pm 0.3$ m s$^{-1}$ with four harmonics. While the coefficients and phases of the sine functions changed with each test, $K$ remained compatible with the value from our adopted model. We also checked that $K$ is not correlated with any other model parameters in the MCMC distribution.

We estimate the probability that RV noise fluctuations conspired to produce an apparently coherent signal with the precise period and phase of Kepler-78b to be approximately one in 16,000. This is the probability of a 4-σ outlier for a normally distributed random variable. We adopt 4-σ because the fractional error on $K$ is approximately 4. Note that this is not the false alarm probability (FAP) commonly computed for new Doppler detections of exoplanets. In those cases one must search over a wide range of orbital periods and phases to detect the planet, and also measure the planet's mass. Here the existence of the planet was already well established[6]. Our job was to measure the planet's mass given knowledge of its orbit.

**Offset-slope RV Spot Model.** To gauge the sensitivity of our results to model assumptions, we considered a second RV model. Like the harmonic spot model, the offset-slope model consists of two components. The Doppler signal from the planet is a sinusoidal function of time with the period and phase held fixed at the values from the photometric analysis. The spot variations are approximated as linear functions of time with slopes and offsets specific to each night, providing much greater model flexibility[43]. This model for the RV at time $t$ on night $n$ is

$$RV(t) = -K \sin(2\pi(t-t_c)/P) + \gamma_n + \dot{\gamma}_n(t - t_n),$$

where $\gamma_n$ is an RV offset, $\dot{\gamma}_n$ is an RV slope (velocity per unit time), and $t_n$ is the median time of observation specific to night $n$. The other symbols have the same meanings as above. As with the previous model, an RV jitter term was added in quadrature to the errors and $P$ and $t_c$ were fixed (Table 1). Altogether, the offset-slope model contains 18 free parameters.

We used an MCMC algorithm to explore the model parameter space. The best-fit model and randomly selected models from the MCMC chain are shown in Extended Data Figure 3. The key result is $K = 1.53 \pm 0.45$ m s$^{-1}$, which is consistent with the value from the harmonic spot model. The lower precision of the offset-slope model (3.4-$\sigma$ vs. 4.1-$\sigma$ significance) results from greater model flexibility. The slopes and offsets on nearby nights are not constrained to produce continuous spot variations as a function of time. For this reason we adopt the harmonic spot model.

As an additional test of the sensitivity to model details, we used the offset-slope framework to model a subset of the RVs. Within each night, we selected the median values from each group of three RVs ordered in time. This selection naturally rejects outlier RVs and matches the observing style on nights 2-8 when three groups of three measurements were made as close as possible to orbital quadratures (maximum or minimum RV). (On night 8, the final group of RVs only has two measurements; we used the mean of those two RVs for this test.) Our MCMC analysis of these median RVs gave a similar result, $K = 1.26 \pm 0.38$ m s$^{-1}$, that is consistent with the above results at the ~1-$\sigma$ level. We conclude that our detection of Kepler-78 is not strongly sensitive to spot model assumptions or to individual RV measurements.

**Additional References:**

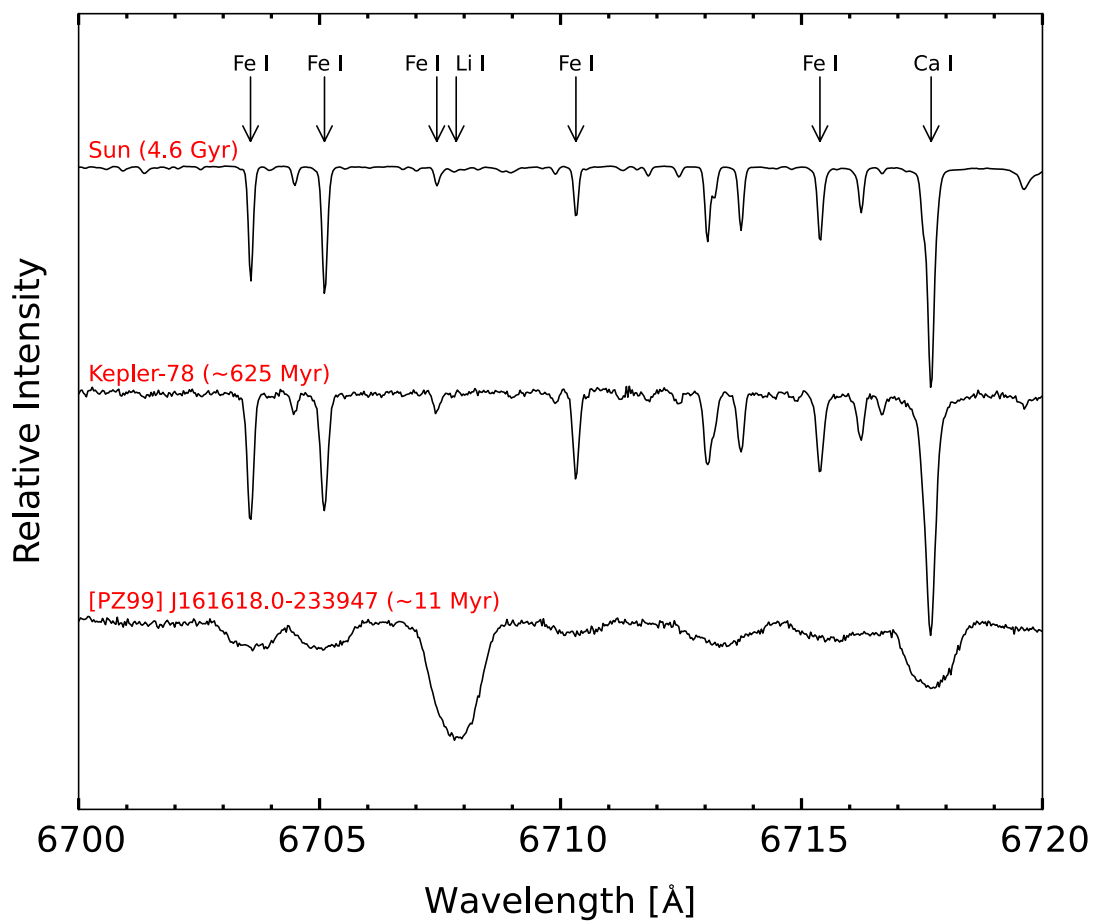

**Extended Data Figure 1 | Wavelength-calibrated spectra of three stars near the age-sensitive Li I line (6708 Å).** This line is not detected in the Kepler-78 spectrum suggesting that Li has been depleted, consistent with an age > 0.5 billion years for this K0 star. The lithium line is also not detected in the 4.6 billion year old Sun. It is clearly seen in the rotationally broadened spectrum of [PZ99] J161618.0-233947, a star whose spectral type (G8) is similar to Kepler-78, but that is much younger (~11 million years)[44]. Additional iron and calcium lines are labeled.

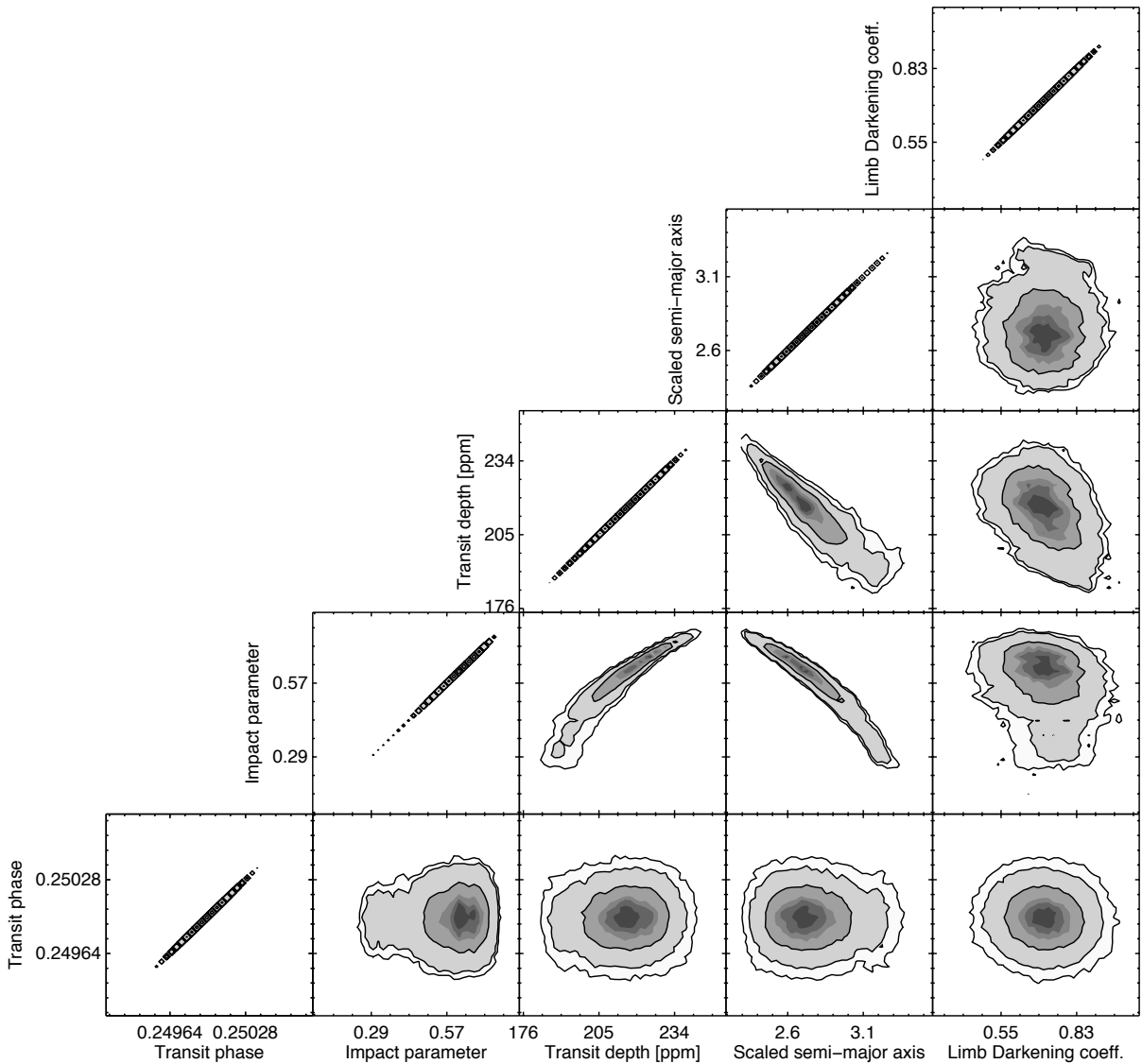

**Extended Data Figure 2 | Correlations between model parameters in the transit analysis.** Grayscale contours denote confidence levels with thick black lines highlighting the 1-σ, 2-σ, and 3-σ contour levels. The strongest correlations are between transit depth, scaled semi-major axis, and impact parameter.

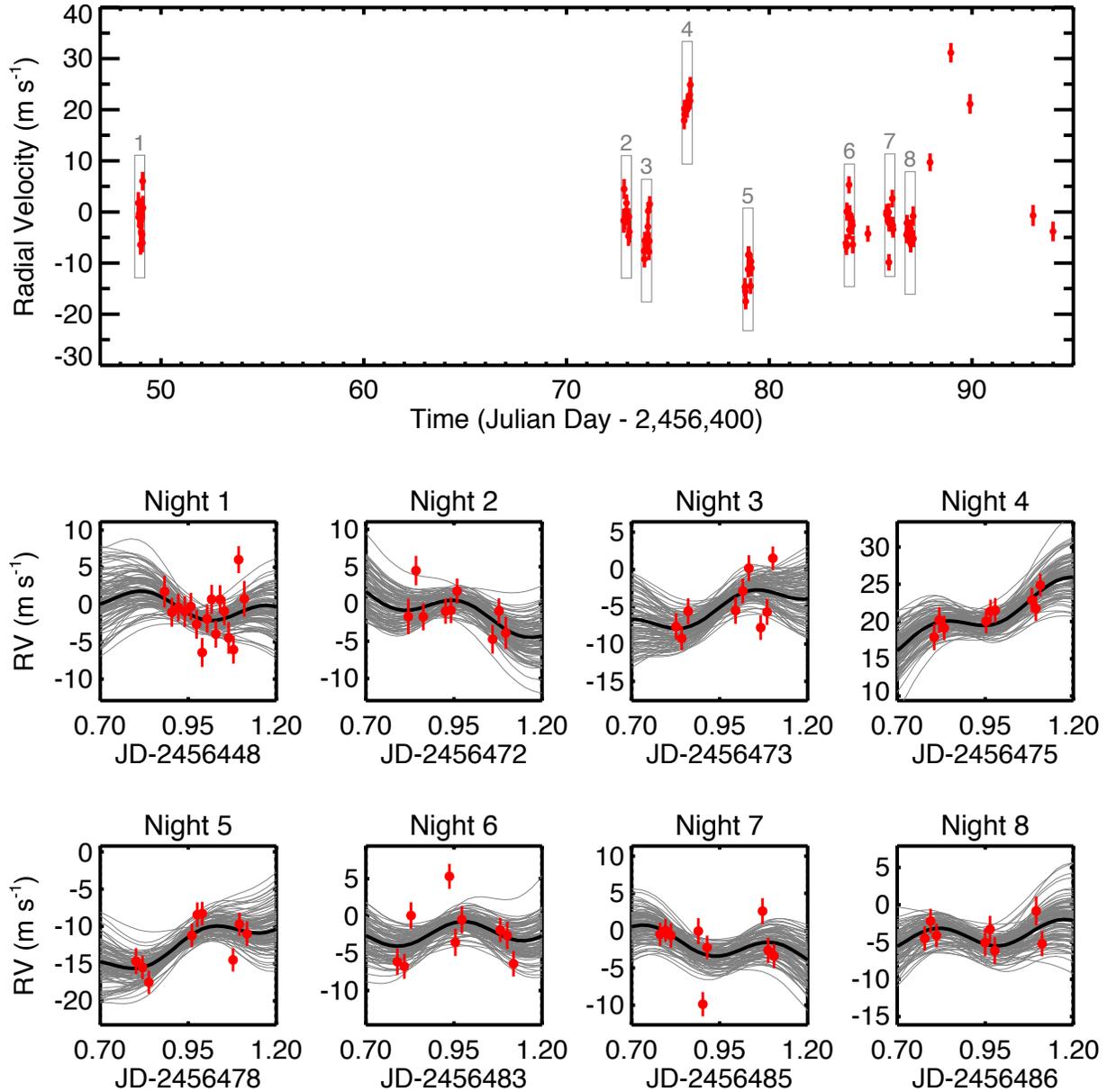

**Extended Data Figure 3 | Apparent radial-velocity (RV) variations of Kepler-78 for the offset-slope model.** The top panel shows the complete 38-day time series of relative RVs (red filled circles). Eight gray boxes highlight nights with intensive observations. The measurements from these nights are shown in the eight subpanels. In each subpanel, the RVs (red filled circles) and best-fit offset-slope model (solid black line) are represented. The RV curves for 100 randomly selected models from the MCMC chain are underplotted in gray, showing the range of variation within the model distribution.

**Supplementary Table 1**. Time series radial velocity and activity measurements from Keck-HIRES spectra. For one measurement (marked "N/A") the $S_{HK}$ index could not be measured because of a cosmic-ray spike on the CCD in the core of the Ca II H line.

| HJD - 2,456,400 | RV (m s$^{-1}$) | RV error (m s$^{-1}$) | $S_{HK}$ Index |
|---|---|---|---|
| 48.88278 | +1.75 | 2.14 | 0.374 |
| 48.90334 | -1.03 | 1.93 | 0.459 |
| 48.92171 | -0.38 | 1.85 | 0.454 |
| 48.93995 | -0.93 | 2.07 | 0.456 |
| 48.95738 | -0.26 | 1.84 | 0.456 |
| 48.97341 | -2.64 | 1.93 | 0.488 |
| 48.98941 | -6.42 | 1.95 | 0.467 |
| 49.00348 | -1.91 | 1.88 | 0.461 |
| 49.01627 | +0.71 | 1.94 | 0.457 |
| 49.02821 | -3.96 | 1.82 | 0.459 |
| 49.04018 | +0.70 | 1.89 | 0.459 |
| 49.05200 | -0.89 | 1.88 | 0.463 |
| 49.06446 | -4.47 | 2.12 | 0.458 |
| 49.07811 | -6.04 | 1.86 | 0.460 |
| 49.09286 | +6.03 | 1.82 | 0.452 |
| 49.10900 | +0.78 | 2.39 | 0.458 |
| 72.82037 | -1.69 | 2.37 | 0.441 |
| 72.84218 | +4.47 | 1.98 | 0.431 |
| 72.86294 | -1.72 | 1.82 | 0.401 |
| 72.92619 | -0.98 | 1.69 | 0.434 |
| 72.94153 | -0.85 | 1.72 | 0.429 |
| 72.95886 | +1.73 | 1.69 | 0.415 |
| 73.05970 | -4.73 | 1.93 | 0.408 |
| 73.07770 | -0.94 | 1.69 | 0.407 |
| 73.09709 | -3.89 | 2.16 | 0.397 |
| 73.82630 | -7.65 | 1.76 | 0.393 |
| 73.84243 | -9.24 | 1.61 | 0.429 |
| 73.86077 | -5.60 | 1.72 | 0.437 |
| 73.99527 | -5.50 | 1.80 | 0.428 |
| 74.01574 | -2.89 | 1.65 | 0.423 |
| 74.03286 | +0.18 | 1.74 | 0.432 |
| 74.06661 | -7.79 | 1.66 | 0.424 |
| 74.08462 | -5.70 | 1.73 | 0.424 |
| 74.10100 | +1.51 | 1.57 | 0.435 |
| 75.80495 | +17.91 | 1.76 | 0.479 |
| 75.81937 | +20.18 | 1.76 | 0.462 |
| 75.83325 | +19.13 | 1.63 | 0.472 |
| 75.95222 | +20.03 | 1.63 | 0.468 |
| 75.96478 | +21.38 | 1.55 | 0.461 |
| 75.97800 | +21.50 | 1.65 | 0.460 |
| 76.08215 | +22.88 | 1.63 | 0.435 |
| 76.09379 | +21.74 | 1.65 | 0.436 |
| 76.10619 | +24.88 | 1.51 | 0.427 |
| 78.80182 | -14.68 | 1.77 | 0.460 |
| 78.82010 | -15.51 | 1.57 | 0.486 |
| 78.83807 | -17.50 | 1.56 | 0.466 |
| 78.96021 | -11.22 | 1.59 | 0.480 |
| 78.97544 | -8.44 | 1.65 | 0.464 |
| 78.98947 | -8.28 | 1.57 | 0.473 |
| 79.07694 | -14.50 | 1.56 | 0.458 |
| 79.09508 | -9.71 | 1.57 | 0.419 |
| 79.11666 | -10.99 | 1.63 | N/A |
| 83.78932 | -6.14 | 1.77 | 0.428 |
| 83.80904 | -6.75 | 1.66 | 0.435 |
| 83.82822 | +0.06 | 1.77 | 0.420 |
| 83.93703 | +5.31 | 1.68 | 0.424 |
| 83.95393 | -3.53 | 1.83 | 0.417 |
| 83.97240 | -0.51 | 1.82 | 0.424 |
| 84.08139 | -1.89 | 1.62 | 0.405 |
| 84.10126 | -2.62 | 1.83 | 0.398 |
| 84.11910 | -6.41 | 1.72 | 0.396 |
| 84.86704 | -4.26 | 1.59 | 0.408 |
| 85.78098 | -0.48 | 1.54 | 0.422 |
| 85.79643 | +0.04 | 1.54 | 0.423 |
| 85.81186 | -0.64 | 1.64 | 0.402 |
| 85.88961 | -0.07 | 1.75 | 0.409 |
| 85.90234 | -9.86 | 1.63 | 0.421 |
| 85.91463 | -2.25 | 1.59 | 0.411 |
| 86.07175 | +2.62 | 1.74 | 0.430 |
| 86.08761 | -2.65 | 1.69 | 0.407 |
| 86.10455 | -3.40 | 1.64 | 0.409 |
| 86.77816 | -4.51 | 1.63 | 0.415 |
| 86.79474 | -2.18 | 1.64 | 0.434 |
| 86.81158 | -4.12 | 1.59 | 0.437 |
| 86.94956 | -5.07 | 1.84 | 0.437 |
| 86.96350 | -3.30 | 1.81 | 0.438 |
| 86.97688 | -6.16 | 1.80 | 0.440 |
| 87.09450 | -0.84 | 1.93 | 0.415 |
| 87.11122 | -5.23 | 1.68 | 0.404 |
| 87.93100 | +9.70 | 1.75 | 0.442 |
| 88.95482 | +31.16 | 1.91 | 0.496 |
| 89.90345 | +21.13 | 1.93 | 0.474 |
| 93.01254 | -0.70 | 2.07 | 0.437 |
| 93.99643 | -3.82 | 1.93 | 0.429 |